\documentclass[fleqn,twoside]{article}
\usepackage{espcrc2}

\usepackage{epsfig}

\newcommand{\AmS}{{\protect\the\textfont2
  A\kern-.1667em\lower.5ex\hbox{M}\kern-.125emS}}

\title{Lattice QCD at finite isospin chemical potential and temperature.}

\author{J.~B.~Kogut\address{Physics Department, University of Illinois, 
                            1110 West Green Street, Urbana, IL 61801, USA}
   and  D.~K.~Sinclair\address{HEP Division, Argonne National Laboratory,
                               9700 South Cass Avenue, Argonne, IL 60439, USA}
\thanks{Talk presented by D.~K.~Sinclair. DKS was supported by DOE contract
W-31-109-ENG-38. JBK was supported in part by an NSF grant NSF PHY-0102409. 
Simulations used IBM SPs and CRAYs at NERSC and NPACI.}}

\begin{document}
\begin{abstract}
We simulate lattice QCD at a finite chemical potential $\mu_I$ for isospin
($I_3$) at zero and finite temperatures. At some $\mu_I=\mu_c$ QCD has a
second order transition with mean-field critical exponents to a state where 
($I_3$) is broken spontaneously by a charged pion condensate. Heating the
system with $\mu_I > \mu_c$ we find there is some temperature at which this
condensate evaporates. This transition appears to be second order and mean-field
at lower $\mu_I$ values, and first order for $\mu_I$ sufficiently large. We
are determining the dependence of the finite temperature crossover $T_c$
on $\mu_I$ for $\mu_I < \mu_c$. This is expected to be identical to $T_c$'s
dependence on quark-number chemical potential $\mu_q$ for small $\mu_q$.  
\end{abstract}

\maketitle

\section{Introduction}

QCD at finite baryon-number density describes nuclear matter including neutron
stars and heavy nuclei. Nuclear matter at finite temperature may be observed 
in relativistic heavy-ion collisions at CERN and RHIC.

Because of the sign/phase difficulties associated with simulating QCD at finite
baryon-number chemical potential, we have been studying related theories with
positive fermion determinants. In particular, we are studying QCD (2-flavours)
at finite chemical potential $\mu_I$ for isospin ($I_3$), at zero and finite
temperature $T$. This represents a surface in the phase diagram for nuclear
matter, which exists at finite isospin ($I_3$) density as well as finite
baryon-number density.

From effective field theory (chiral perturbation theory) and other analyses,
we expect a $T=0$ transition at $\mu_I=m_\pi$ to a state with a charged pion
condensate which breaks $I_3$ spontaneously, with the orthogonal charged pion
state becoming a Goldstone boson \cite{sonstep} The transition is expected to
be second order with mean-field critical exponents \cite{sonstep,stv}, a result
which our simulations confirm \cite{isospin}.

For $\mu_I > m_\pi$ there is a finite $T$ phase transition where the pion
condensate evaporates. For $\mu_I$ large enough, this transition appears to
be first order. At low $T$ the transition at $\mu_I=\mu_c \ge m_\pi$ appears
to be second order and mean field as it is for $T=0$ \cite{isospin}.

At $\mu_I=0$ ($m \ne 0$) we know that there is a rapid crossover from `hadronic
matter' to a `quark-gluon' plasma at $T=T_c \sim 150-200$MeV. This is the
start of a line of such crossovers, which extends to finite $\mu_I$. For small
$\mu_I$, the Swansea-Bielefeld collaboration have noted that 
$\beta_c(\mu_q) = \beta_c(\mu_I=2 \mu_q)$ to the extent that $\beta_c$ is well
defined for a crossover \cite{swanbiel}.

\section{QCD at finite $\mu_I$}

The lattice quark action is
\begin{equation}
S_f=\sum_{sites} \left[\bar{\chi}[D\!\!\!\!/(\frac{1}{2}\tau_3\mu_I)+m]\chi
                   + i\lambda\epsilon\bar{\chi}\tau_2\chi\right]
\end{equation}
which has a positive fermion determinant. $\lambda$ is the explicit $I_3$ 
breaking parameter, required to observe spontaneous symmetry breaking on a
finite lattice. We are interested in $\lambda \rightarrow 0$.

Our zero temperature simulations were performed on an $8^4$ lattice at
$m=0.025$ and $m=0.05$ at $\lambda=0.1 \times m$ and $\lambda=0.2 \times m$,
using the hybrid molecular-dynamics method with `noisy' fermions, and tuning to
$N_f=2$.

\begin{figure}[htb]
\epsfxsize=3in 
\centerline{\epsffile{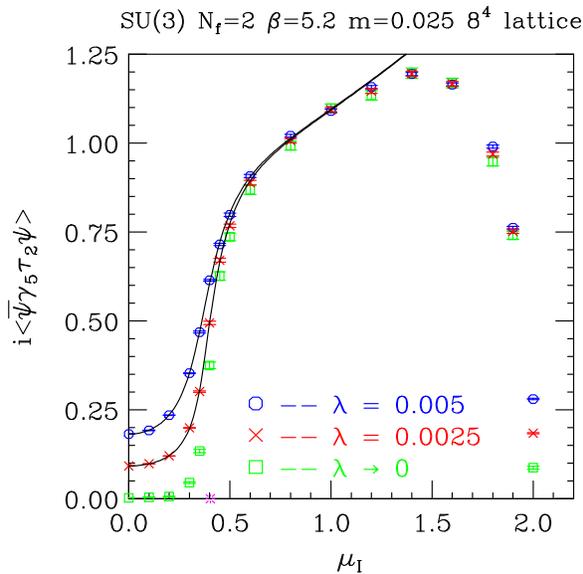}}
\caption{Charged pion condensate as a function of $I_3$ chemical
potential $\mu_I$. The curves are a fit to mean-field scaling.}%
\label{fig:zeroT}
\end{figure}

Figure \ref{fig:zeroT} shows the charged pion condensate for the $m=0.025$ runs,
which shows evidence for a phase transition at $\mu_I=\mu_c$ to a state with
a charged pion condensate at $\lambda=0$. The fit is to a mean-field scaling
form suggested by effective Lagrangian analyses.

For more details and measurements of other observables we refer the reader to
\cite{isospin}.

\section{QCD at finite $\mu_I$ and T}
 
We have performed simulations on $8^3 \times 4$ lattices to study the finite
temperature ($T$) behaviour of lattice QCD at finite $\mu_I$ \cite{isospin}.

Simulations at low $\beta$ ($4.0$ and $5.0$) and hence $T$ have been performed,
varying $\mu_I$. Again we find that there is a phase transition for $\lambda=0$
at some $\mu_c(\beta)$, above which the system is in a phase characterized by
a charged pion condensate which breaks $I_3$ spontaneously. The transition
appears to be second order with mean-field critical exponents, as in the $T=0$
case.

We have also performed simulations at fixed $\mu_I > \mu_c(T=0)$, increasing
$\beta$ (and hence $T$) until the pion condensate `evaporates' and the $I_3$
symmetry is restored. For large $\mu_I$ this phase transition is sufficiently
abrupt to suggest that the transition has become first order. We are currently
repeating these simulations on a larger ($16^3 \times 4$) lattice to check
that the transition is indeed first order.

\begin{figure}[htb]    
\epsfxsize=3in                                                             
\centerline{\epsffile{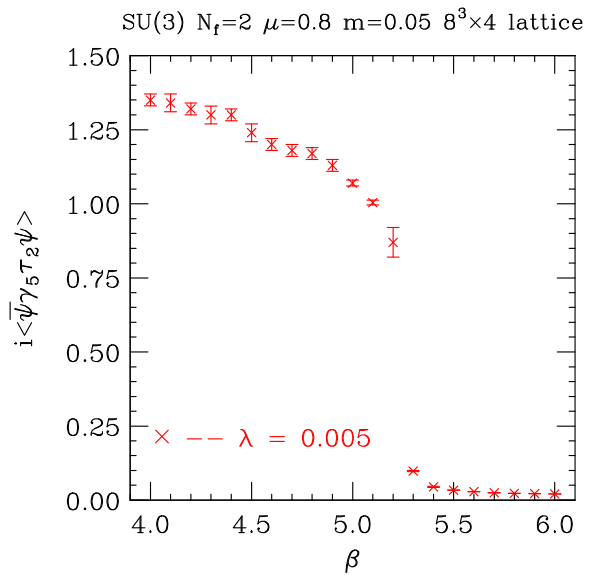}}
\caption{Pion condensate as a function of $\beta$ for $m=0.05$,
$\lambda=0.005$, and $\mu_I=0.8$ on an $8^3 \times 4$ lattice.}%
\label{fig:pi_mu8}                                              
\end{figure}                                                    
  
Figure~\ref{fig:pi_mu8} shows the pion condensate for $\mu_I=0.8$ as a function
of $\beta$. The abrupt `jump' between $\beta=5.2$ and $\beta=5.3$ suggests a
first order transition. Preliminary simulations on a $16^3 \times 4$ lattice
indicate that there is such an abrupt transition between $\beta=5.265$ and
$\beta=5.270$.

\section{The finite T transition at small $\mu_I$}

The finite temperature crossover ($m \ne 0$) from hadronic matter to a
quark-gluon plasma extends to finite $\mu_I$. We are running on $8^3 \times 4$
lattices with $m=0.05$ and $N_f=2$. For $\mu_I < \mu_c(T=0)=0.569(2)$ we set
$\lambda=0$.

At small $\mu_q$, the Swansea-Bielefeld collaboration \cite{swanbiel} found
that the phase $\theta$ of the fermion determinant was `well-behaved' --
$\langle\cos(\theta)\rangle > 0$ and goes smoothly to $1$ as $\mu_q
\rightarrow 0$. As a consequence they expect, and find numerically by their
series expansion methods, that $\beta_c(\mu_q) = \beta_c(\mu_I=2 \mu_q)$ to
the extent that $\beta_c$ is well defined for a crossover. Thus for small
$\mu_I$ our simulations predict the small $\mu_q$ behaviour of $\beta_c$.

We have performed simulations at $\mu_I=0.0,0.1,0.2,0.3,0.4$, and are currently
running at $\mu_I=0.5$, varying $\beta$ through the crossover regime, measuring
the chiral condensate, Wilson line, plaquette, isospin density and their
susceptibilities, from which we will obtain the position of the crossover (the
maxima of the susceptibilities) as a function of $\mu_I$.

\begin{figure}[htb]
\epsfxsize=3in
\centerline{\epsffile{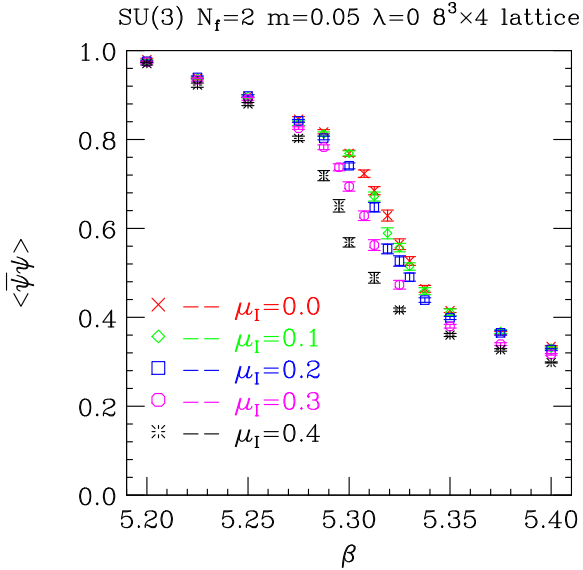}}
\caption{Chiral condensates for $\mu_I=0.0,0.1,0.2,0.3,0.4$, as functions
of $\beta$.}%
\label{fig:chiral}
\end{figure}

Figure~\ref{fig:chiral} shows the chiral condensate measurements from these
runs. We note that the temperature of the transition decreases {\it slowly} 
with increasing $\mu_I$ in qualitative agreement with the $\mu_q$ behaviour
reported by others \cite{fk,swanbiel,defp,lattice2002}.

\section{Conclusions}

At zero and low temperatures, there is some finite $\mu_I$, $\mu_c$ say, above
which $I_3$ is spontaneously broken by a charged pion condensate. This phase
transition appears to be second order with mean-field critical exponents.
Simulations will be performed on a $12^3 \times 24$ lattice to measure the
spectrum of Goldstone and pseudo-Goldstone bosons, and to study instanton
effects.

For $\mu_I > \mu_c$ there is a finite temperature phase transition at which
the pion condensate evaporates. For large $\mu_I$ this transition appears to
be first order. Simulations on a $16^3 \times 4$ lattice are being performed
to check this.

We are studying the crossover from hadronic matter to a quark-gluon plasma at
small $\mu_I$ and have confirmed that $\beta_c$ and hence $T_c$ decrease very
slowly with increasing $\mu_I$. This observation will be made quantitative.
This is made more interesting because at low $\mu_I$, the $\mu_I$ and $\mu_q$
dependence of this transition are believed to be identical. Simulations are
planned to extend these studies of the $\mu_I$ dependence of the finite
temperature transition to intermediate $\mu_I$ values. Larger-lattice
simulations are needed.

It would be interesting to know if such charged pion condensates form when the
system is also at finite baryon number density.

\end{document}